\documentclass[12pt]{article}
\usepackage{hyperref}
\usepackage{amsmath}
\usepackage{graphicx}
\usepackage[font=small]{caption}
\usepackage[affil-it]{authblk}
\advance \topmargin by -\headheight
\advance \topmargin by -\headsep   
\evensidemargin \oddsidemargin
\newcommand{\beq}{\begin{equation}}
\newcommand{\eeq}{\end{equation}}
\newcommand{\mcal}{\mathcal}

\begin{document}

\title{Modeling Market Inefficiencies within a Single Instrument}
\author{Kuang-Ting Chen}
\affil{Deutsche Bank, New York, NY 10005}
\maketitle
\abstract{In this paper, we propose a minimal model beyond geometric Brownian motion that aims to describe price actions with market inefficiency. From simple financial theory considerations, we arrive at a simple two-variable hidden Markovian time series model, with one of the variable entirely unobserved. Then, we analyze the simplest version of the model, using path integral and Green's function techniques from physics. We show that in this model, the inefficient market price is trend-following when the standard deviation of the log reasonable price ($\sigma$) is larger than that of the log market price ($\sigma'$), and mean-reversing when it is smaller. The risk premium is proportional to the difference between the current market price and the exponential moving average (EMA) of the past prices. This model thus provides a theoretical explanation how the EMA of the past price can directly affect future prices, i.e., the so-called ``Bollinger bands" in technical analyses. We then carry out a maximum likelihood estimate for the model parameters from the observed market price, by integrating out the reasonable price in Fourier space. Finally we analyze recent S\&P500 index data and see to what extent the real world data can be described by this simple model.}
\thispagestyle{empty}
\newpage
\section{Introduction}
In financial theory, assets are priced according to the expected discounted payoff\cite{nobel2013}:
\beq\label{p}
P=E(Mx).
\eeq
The discount factor $M$ describes both how future money is less valuable due to both the interest rate, and more importantly, how people value various outcomes differently. Take stock and insurance as example: in the United States, the stock market has an annual return of 8\% in the long term, which is significantly higher than the US treasury bond rate at at most 2\%-3\% a year. The interpretation of this fact is that people are worried about losing money potentially; therefore when they price assets such as stocks that may lose value, they discount more when they have a profit, and less when they have a loss. This results in the fact that the price of the stock is lower than the discounted expected payoff. (Notice the opposite order of words; price is the expected discounted payoff where one take the expectation last, whereas the naive valuation process we usually have in mind is to take the expectation at the future where the payoff occurs, then discount it back to the present.) In contrast, when one buys insurance, one is paying extra for that in case the unfortunate happens, he has enough payoff to deal with it. In this case one put more weight on the unfortunate scenario; the price thus ends up higher than the discounted expected payoff. In theory, the prices are therefore not necessarily lower or higher than the discounted expectation, but depends on how people value the various outcome. The difference between price and the discount expected payoff, divided by the standard deviation of the instrument, is usually called the risk premium. So we say that stocks usually have positive risk premium, and insurances have negative risk premium.

If the asset in question is freely tradable, the price is then determined by the market. In particular, the supply, i.e., number of people willing to sell the asset at the market price, should meet the demand, the number of people willing to buy at market price. The discount factor which prices the asset in reality, thus describes the people who are ambivalent toward buying or selling the asset at the market price. In this sense, when there is a market price, we can always derive an effective discount factor, given the forecast, and the current market price.

What is the idea of efficient market then? The usual statement of the efficient market hypothesis (EMH) is that stocks are always traded at fair value, and their value already reflect all available information. A detailed discussion can be found in Ref. \cite{nobel2013} and the references therein. In our words, it means that the price should not deviate significantly from how a \textit{reasonable} discount factor would price it. There are no clear-cut definition of reasonable unfortunately\footnote{This corresponds to the ``joint hyphothesis" problem mentioned in Ref. \cite{nobel2013}.}, but for example, similar instruments should have similar risk premiums; the risk premium should not be too high for any instrument; for one instrument, the risk premium should not vary significantly over time, without any new information. (An important side note is that not all instruments need to have similar risk premium for the market to be efficient. Investors can always benefit from diversifying their portfolio from more independent assets, so instruments with lower risk premium are still valuable, it is just that a reasonable investor will require less quantity of it.) 

The EMH in the weak sense, however, is more concrete. It states that one can never predict returns by analyzing past data. In other words, there are no exploitable opportunities to trade for a larger profit. A lot of work (see for example \cite{cl,ito} and references therein) has been done to address whether such hypothesis is true on various assets, but the results are mixed. There are also efforts made\cite{zunino,zunino2,apen} to develop different statistics to measure market inefficiencies. On the other front, there are also countless time series models that have been proposed\cite{taylor,tim,philip} to fit the observed financial time series, whether satisfying EMH or not. 

In this paper, we take a different approach. Instead of proposing time series models that have a number parameters that can fit the data such as ARIMA or GARCH, we start from the financial theory side. We think about how market efficiency can be broken in a minimal way. When the market is not efficient, we propose a model to replace the geometric Brownian motion as the next simplest description. In the following, in Sec. 2 we start from simple financial theory considerations and motivate the model. In Sec. 3 we first solve the model analytically, and find its current risk premium based on past history. Then we discuss a maximum likelihood procedure to estimate the model parameters from data. In Sec. 4, we take a look at real data and see if such modeled inefficiencies are indeed present. 

\section{A Simple Theory to Model Market Inefficiencies}
In the literature, the simplest model of the price of a stock follows the geometric Brownian motion:
\beq
\frac{dS}{S}=\mu dt+\sigma dz.
\eeq
If we assume that the stock is always priced fairly, then the stock price in the immediate future is the payoff. We can then use the pricing equation relating the immediate price in the future and the current price, via the discount factor:
\beq
S=E\big((1+\frac{d\Lambda}{\Lambda})(S+dS)\big).
\eeq
For simplicity if we further assume that the risk-free interest rate is zero, i.e., $E(d\Lambda/\Lambda)=0$, this will imply that the stochastic discount factor is
\beq
\frac{d\Lambda}{\Lambda}=-\frac{\mu}{\sigma}dz.
\eeq
This discount factor can then be used to price derivatives such as options.

A few observations:

1. The discount factor relates the current price to the price in the future. This is because when the pricing is fair, the future price is just the discounted payoff at that time.  

2. We derived the discount factor from the random process that we assume the price to follow. This is always possible for a given random process, but there is no guarantee that the resulting discount factor is reasonable. Conversely, we can also presume a discount factor, then deduce how the stock price should behave. The discount factor does not uniquely determine the stock price however; it only determines the ratio between the drift and the standard deviation.

To model an inefficient market, the simplest way is to keep the pricing equation intact, but forgo the constraint that the discount factor should be reasonable. However, the random process that governs the prices then depends entirely on how we want to choose this unreasonable discount factor. Therefore, the real question is always how to reasonably define a unreasonable discount factor.

\subsection{the Fundamental Problem with Market Inefficiency}
One way of choosing this effective unreasonable discount factor is to assume a reasonable discount factor, but then allow the price to deviate from the pricing equation, Eq. \ref{p}. That is, we imagine there is a reasonable price following the pricing equation of the reasonable discount factor, but the real price does not always match the reasonable price. In terms of equations, it roughly looks like the following:
\beq\label{p0}
P_0=E(Mx);
\eeq
\beq\label{rp}
\frac{dP}{P}=f(P,P_0)dt+\sigma dz'.
\eeq
$P_0$ is the reasonable price, $M$ is the reasonable discount factor we have chosen, $x$ is the payoff, and $P$ is the real price. The second equation is necessary; since we have decoupled the real price from the pricing equation, we need to give additional information on how it is going to evolve. $f$ would be some function that makes the price $P$ track the reasonable price $P_0$ over longer time scales. $dz'$ is a zero mean random noise, different from the random noise in the reasonable price.

This set of equations is not self-consistent, however. The central problem is that if the instrument in question is freely tradable at price $P(t=0)$ at $t=0$, that price constitutes an immediate payoff $x(t=0)$, for which the seller can choose to take if it is higher than his estimate of the long term discounted payoff, and the buyers can choose to pay if it is lower than his estimate. Therefore, the reasonable price $P_0$, being the expected discount payoff, has to be higher than $P$ for the seller, and lower than $P$ for the buyer, if it were to follow Eq. \ref{p0}. For a reasonable price definition independent of the market position, the only possible take is that $P_0=P$, where supply meets demand. 

To be clear, let us consider a simple example. Suppose that the reasonable discount factor $M=1$, and imagine we currently have an``underpriced" stock, whose price/payoff $x$ at a later time is expected to be higher than the current price $P$. How do we define $P_0$? From Eq. \ref{p0}, the maximal payoff for a potential buyer would be $P_0=x$, higher than $P$. A potential seller, on the other hand, can only get his best payoff by buying back immediately, and his reasonable price would be $P_0=P$. This reasonable price, defined by the expected discount payoff, is different for the two sides, because one always has the option to trade whenever he wants. 

Conceptually the reasonable price we want to define in this example is apparently the buyer's, but as we have seen from above, we cannot define it as the immediate expected discount payoff. The ability to trade at market price at any time prohibits the expected discount payoff to deviate from the real price. 

\subsection{Remedy to the Problem}

In order to work around this pitfall, when we define the reasonable price via Eq. \ref{p0}, we have to imagine that the payoff is more about the long term, and the current market price should not have a direct influence. In other words, when defining the payoff $x$ in our theory we have to let it deviate freely from the market price $P$, at least in the short term.

How do we capture the long term payoff using variables at present? The answer is that it is captured by $P_0$ at the very next moment. Indeed, if $P_0$ is the expected discounted payoff at any given time, every $P_0$ captures the long term payoff after its time, and the pricing equation relates $P_0$ at this and the very next moment. Given
\beq
\frac{d\Lambda }{\Lambda}=-rdt-mdz
\eeq 
for example, and if we assume the fluctuation of the reasonable price is proportional to its current value, we can then get
\beq
\frac{dP_0}{P_0}=(r+m\sigma_0)dt+\sigma_0 dz.
\eeq
Here $r$ is the risk-free interest rate, $m$ is the market price of risk, and $\sigma_0$ is the standard deviation of the price percentage of the instrument. $dz$ describes the random incoming shocks that affect the instrument; for example this may include the profitability of a company stock, overall macroeconomic conditions for a treasury bond, or international relations for energy futures.

The real price still evolves according to Eq. \ref{rp}. Here we give further specification to the random noise: both real incoming news and current market force fluctuations should affect the market price. We therefore write the following instead:
\beq
\frac{dP}{P}=f(P,P_0)dt+\sigma_1 dz'+\sigma_2 dz.
\eeq 
$dz'$ represents market force flucuations, and $dz$ is the same shock as in the pricing equation.

In this approach, the reasonable price $P_0$ is now seemingly defined entirely independent of the real market price $P$, whereas the real market price is trying to regress to $P_0$ with some additional noise.  However, current market price can affect the reasonable pricing in some way. The simplest way is to add a term $\sigma_3 dz'$ to the stochastic equation of $P_0$. With this term $P_0$ still satisfies the same pricing equation, since the additional noise $dz'$ does not carry a risk premium. We end up with this following set of equations:

\beq\label{p02}
\frac{dP_0}{P_0}=(r+m\sigma_0)dt+\sigma_0 dz+\sigma_3 dz';
\eeq
\beq\label{rp2}
\frac{dP}{P}=f(P,P_0)dt+\sigma_1 dz'+\sigma_2 dz.
\eeq 
Interestingly though, we cannot really tell apart between $dz$ and $dz$' with just the observation of the two prices. All we can say is how large the fluctuation of each price is, and how they are correlated. If we can deduce $\sigma_0$ from our prior knowledge of the risk-free rate $r$ and risk premium $m$, then we can get to know about all the $\sigma$'s. However, if not, then all possible values of $\sigma$'s which give the same variances of the two prices as well as the correlation between them are equivalent. Therefore, without any prior knowledge of the parameters of the model, the model is equivalent to the following:
\beq\label{p03}
\frac{dP_0}{P_0}=adt+\sigma dz;
\eeq
\beq\label{rp3}
\frac{dP}{P}=f(P,P_0)dt+\sigma' dz';
\eeq
\beq
\mathrm{Cov}(dz,dz')=\rho dt.
\eeq
That is, adding this $\sigma_3dz'$ term into the evolution of $P_0$ actually does not change the dynamics of the model. It just maps the original model to a different set of parameter values. In the remaining of the paper, we will study the dynamics of this set of equations.

Incidentally, the mathematics governing this set of equations is identical to quantum mechanics in Euclidean time. Specifically if we choose $f=-k\ln(P/P_0)$, the theory becomes non-interacting and exactly solvable. We shall stick with this choice in the remaining of the paper.

\section{Analytical Solution of the Model}
Usually when one speaks about the solution of such a set of stochastic differential equations, it refers to a probability distribution as a function of time, given the initial values of the prices. In our case, there are a few important differences: (i) we do not observe the reasonable price $P_0$; (ii) we are fine with just knowing about the probabilities in the immediate future. 

While the time evolution of $P_0$ is simply geometric Brownian motion, to plug in that distribution into the evolution of $X$ and integrate for a finite amount of time, with $P$ itself also at the right hand side of the equation, is no simple task. Fortunately this prediction of finite time horizon is not that important to us as any newly observed price $X$ will change it. It suffices for us to know how $X$ is going to behave only for the next time step. The important question for us is to find the unobserved $P_0$ at the moment, based on the evolution of $P(t)$ in the past. In addition, we would like to infer the model parameters from the observed price as well. 

In the first subsection, we shall assume knowledge of the model parameters, and find the most probable values of $P_0$ as a function of time. In the second subsection, we highlight the result we get for $\ln P_0$ and discuss the resulting dynamics of the market price $P$, in various parameter ranges. In the last subsection, we shall find the maximum likelihood estimate (MLE) of the model parameters, by integrating over all possible values of $P_0$ in Fourier space.

\subsection{Maximum likelihood estimate of $P_0$}
Let $X\equiv\ln P$, $X_0\equiv\ln P_0$, by Ito's Lemma, we then have
\beq
d X_0=(a-\frac{\sigma^2}{2})dt+\sigma dz;
\eeq
\beq
d X=(-k(X-X_0)-\frac{\sigma'^2}{2})dt+\sigma' dz';
\eeq
We can write down the log likelihood function $\ln L(a,k,\sigma,\sigma',\rho;X_0(t),X(t))$:
\beq
\ln L=-\frac{1}{2(1-\rho^2)\sigma^2\sigma'^2}\int dt\bigg(\sigma'^2\Delta X_0^2+\sigma^2\Delta X^2-2\rho\sigma\sigma'\Delta X_0\Delta X\bigg)+C,
\eeq
with
\beq
\Delta X_0\equiv \dot X_0-(a-\sigma^2/2);
\eeq
\beq
\Delta X\equiv \dot X+k(X-X_0)+\sigma'^2/2.
\eeq
C is some function of the parameters that normalizes the likelihood function, and is not a functional of $X$ or $X_0$. Our first step is to find the most probable $X_0$, given the parameters and some history of $X(t)$. The variation on $X_0$ will give the equation that $X_0$ should satisfy to maximize $\ln L$, which is just the equation of motion of $X_0$ in Euclidean time:
\beq
-2\sigma'^2\ddot X_0-2\sigma^2 k\big(\dot X+k(X-X_0)+\sigma'^2/2\big)
+2\rho\sigma\sigma'\big(-k(a-\sigma^2/2)+\ddot X+k\dot X\big)=0;
\eeq
Collecting terms, we get
\beq
X_0-\frac{\sigma'^2}{k^2\sigma^2}\ddot X_0=X+\frac{1}{k}(1-\frac{\rho\sigma'}{\sigma})\dot X-\frac{\rho\sigma'}{k^2\sigma}\ddot X+\frac{\sigma'^2}{2k}+\frac{\rho\sigma'}{k\sigma}(a-\frac{\sigma^2}{2})\equiv g(t).
\eeq
The right-hand side is a known function of time. Let us denote that as $g(t)$, then the most probable $X_0(t)$ shall be
\beq\label{x0sol}
X_0(t)=\frac{k\sigma}{2\sigma'}\int^\infty_{-\infty} dt'g(t')\exp\big(-\frac{k\sigma}{\sigma'}|t-t'|\big).
\eeq 

We can have some intuition about the solution of $X_0$. First, $\sigma'/k\sigma$
defines a time scale at which we can recover $X_0$ from $X$. it is inversely proportional to $k$, because $k^{-1}$ is the characteristic time scale for $X$ to respond to $X_0$. The $\sigma's$ are there because they tell us how "noisy" the prices are; for example, if $\sigma$ is small, then we know that $X_0$ is not fluctuating around much, so we can afford to average longer to get a more accurate estimate. If $\sigma'$ is small, then $X$ carries less noise, so it makes sense to average at a smaller time scale to more timely reflect the change of $X_0$.



There is one important thing we have overlooked up to this point. Notice that the kernel $\frac{k\sigma}{2\sigma'}\exp\big(-\frac{k\sigma}{\sigma'}|t-t'|\big)$ is nonzero even when $t'>t$; that is, the estimate of $X_0$ receives contribution from the prices both before and after it. In physics terminology, we are using the Feynman prescription of the Green's function. This naturally occurs in a maximum likelihood estimate, since the current reasonable price evolves gradually from the past reasonable prices (so that it is influenced by the previous prices) and it affects the future prices (so that we need to draw inferences from the future prices to make the estimate.)  Furthermore, mathematically this is the only converging Green's function. Specifically, if we think of the problem as an initial value problem (in other words, if we are looking for a retarded Green's function), the solution for a generic $g(t)$ is always running away to $\pm\infty$, due to the same sign between $X_0$ and $\ddot X_0$.

The maximum likelihood estimate of $X_0$ in the middle of a time series is therefore not suitable to be used to understand the dynamics of $X$, as it already contains information from the future. What we are interested instead is something like a maximum likelihood estimate without knowledge to the prices afterwards. What we can think about in our formulation, is then the estimate of $X_0$ at the end of the time series. The special solution, Eq. \ref{x0sol}, is not up for the task unfortunately, because we have not specified what boundary condition it should satisfy at the end of the time series. 

The general solution of $X_0$ is in the following form:
\beq\label{gensol}
X_0=\frac{k\sigma}{2\sigma'}\int^T_{0} dt'g(t')\exp\big(-\frac{k\sigma}{\sigma'}|t-t'|\big)+a\exp(-\frac{k\sigma}{\sigma'}t)+b\exp(\frac{k\sigma}{\sigma'}(t-T)),
\eeq
where $a$ and $b$ are constants determined by the boundary conditions. $a$ can be thought of as some prior knowledge about $X_0$ at the start of the time series; for our purpose as long as $(k\sigma /\sigma')T\gg 1$, it does not affect $X_0(T)$. The question is $b$. What boundary condition should we take at the end of the time series in order to determine $b$?

It turns out the answer is simple. When we do the variation to maximize the log likelihood, we have integrated by parts to change terms proportional to $\delta \dot X_0$ to  $\delta X_0$. This results in a total derivative, which we have discarded in our derivation of the equation of motion in the bulk. This total derivative will determine the boundary condition that the solution of $X_0(t)$ needs to follow, as shown below: 

In the variation, we have
\begin{align}
\delta \ln L&=\int^T_0 dt\;\frac{\partial \mcal L}{\partial X_0}\delta X_0+\frac{\partial \mcal L}{\partial \dot X_0}\delta \dot X_0\nonumber\\
&=\int^T_0 dt\;\big(\frac{\partial \mcal L}{\partial X_0}-\frac{\rm d}{{\rm d}t}\frac{\partial \mcal L}{\partial \dot X_0}\big)\delta X_0+\big(\frac{\partial \mcal L}{\partial \dot X_0}\delta X_0\big)|^T_0;
\end{align}
here we borrow the Lagrangian symbol $\mcal L\equiv \big(\sigma'^2\Delta X_0^2+\sigma^2\Delta X^2-2\rho\sigma\sigma'\Delta X_0\Delta X\big)/2\sigma^2\sigma'^2(1-\rho^2)$ to denote the integrand in the log likelihood function. In addition to the terms in the bulk proportional to the equation of motion, we thus still have a total derivative that can be integrated to the boundary,
\begin{align}
\delta \ln L&({\rm boundary})=\big(\frac{\partial \mcal L}{\partial \dot X_0}\delta X_0\big)|^T_0\nonumber\\
&=-\frac{\delta X_0}{(1-\rho^2)\sigma^2}\bigg(\dot X_0-(a-\sigma^2/2)-\frac{\rho\sigma}{\sigma'}\big(\dot X+k(X-X_0)+\sigma'^2/2\big)\bigg)|^T_0.
\end{align}
Setting it to zero, we can get an equation at the boundary which relates $X_0$ to $\dot X_0$:
\beq\label{bc} 
\dot X_0(T)-(a-\sigma^2/2)-\frac{\rho\sigma}{\sigma'}\big(\dot X(T)+k(X(T)-X_0(T))+\sigma'^2/2\big)=0.
\eeq
This is the boundary condition we are looking for. 

There is a neat trick to find the solution of $X_0(t)$. First we extend the range of time to $2T$ and define
\beq 
g(t-T)=g(2T-t); \;T<t<2T.
\eeq
That is, in the extra range $g(t)$ is a mirror image of the original $g(t)$. It is important to notice that the induced definition of $X(t)$ in general is not symmetric around $t=T$. Now we write our solution as
\beq\label{gensol2}
X_0=\frac{k\sigma}{2\sigma'}\int^{2T}_{0} dt'g(t')\exp\big(-\frac{k\sigma}{\sigma'}|t-t'|\big)+b'\exp(\frac{k\sigma}{\sigma'}(t-T)).
\eeq
This is just a way to rewrite the same general solution, as we can see if we integrate out $t'$ in the range $T<t'<2T$, for $0<t<T$ we recover Eq. \ref{gensol}, with
\beq
b=b'+\frac{k\sigma}{2\sigma'}\int^{T}_{0} dt'\;g(t')\exp\big(-\frac{k\sigma}{\sigma'}(T-t')\big).
\eeq
Due to the fact that the first term in Eq. \ref{gensol2} is an even function under $t\rightarrow 2T-t$, at $t=T$, its time derivative vanishes. We now can write $X_0$ and $\dot X_0$ as
\begin{align}\label{sol}
X_0(T)&=\frac{k\sigma}{\sigma'}\int^{T}_{0} dt'g(t')\exp\big(-\frac{k\sigma}{\sigma'}(T-t')\big)+b'\nonumber\\
&\equiv \bar {X_0}+b';\\
\dot X_0(T)&=\frac{k\sigma}{\sigma'}b'.
\end{align}
Plugging into Eq. \ref{bc}, we get
\beq
b'=\frac{\sigma'}{k\sigma(1+\rho)}\big(a-\sigma^2/2+\frac{\rho\sigma}{\sigma'}(\dot X+k(X-\bar X_0)+\sigma'^2/2)\big).
\eeq

The answer is complicated, but it simplifies at the end of the time series. In fact, one surprising feature is that $X_0(T)$ is actually independent of $\rho$. To see this, let us consider the differential equation that $X_0(T)$ needs to follow (that is, when we gradually accumulate more data and lengthen our time series, how does the most probable $X_0(T)$ at the end of the current time series change.)

First notice that $\bar X_0$ is the only integral that appears in $X_0(T)$, and it satisfies
\beq\label{barx0eq}
\bar X_0(T)+\frac{\sigma'}{k\sigma}\frac{{ d}\bar X_0(T)}{{ d}T}=g(T).
\eeq

The remaining part
\begin{align}
X_0(T)-\frac{\bar X_0(T)}{1+\rho}&=\frac{\sigma'}{k\sigma(1+\rho)}\big(a-\sigma^2/2+\frac{\rho\sigma}{\sigma'}(\dot X+kX+\sigma'^2/2)\big)\nonumber\\&\equiv C(T)
\end{align}
depends only on the local information at $T$. Plugging in $\bar X_0(T)=(1+\rho)(X_0(T)-C(T))$ to Eq. \ref{barx0eq}, we then get
\beq
X_0(T)+\frac{\sigma'}{k\sigma}\dot X_0(T)=\frac{1}{1+\rho}g(T)+C+\frac{\sigma'}{k\sigma}\dot C.
\eeq
Combining the right hand side, we find that the $\rho$ dependence completely cancels out, and the equation becomes
\beq
X_0(T)+\frac{\sigma'}{k\sigma}\dot X_0(T)=\frac{\sigma'}{k\sigma}(a-\sigma^2/2)+\frac{\sigma'^2}{2k}+X(T)+\frac{1}{k}\dot X(T)\equiv h(T).
\eeq
We can also explicitly write the solution as
\beq\label{x0ind}
X_0(T)=\frac{k\sigma}{\sigma'}\int^{T}_{0} dt'h(t')\exp\big(-\frac{k\sigma}{\sigma'}(T-t')\big).
\eeq
An alternative way of deriving the same thing is to integrate by part all the terms proportional to $\rho$ in $g(t)$ in Eq. \ref{sol}. By explicit calculation one will see all $\rho$ dependence cancels. 

What does this mean? Naively it sounds paradoxical. How can the estimate of $X_0$ be independent of $\rho$, which is the correlation between the fluctuation of $X_0$ and the observed $X$? This is because this equation describes strictly the estimate of $X_0$ at the very end of the time series. Once new data comes in, the best estimate at that time needs to be updated. And the updated estimate will have dependence on $\rho$. In other words, for a given time series, the estimate for $X_0(t)$ will be different for different $\rho$, but they will all end up at the same point.

Interestingly, this necessarily means that we can never know about the true $\rho$ from data, if we only observe $X(t)$. This is because the evolution of $X(t)$ depends only on the difference between $X_0$ and $X$. When our estimate of $X_0$ is independent of $\rho$, it implies for any $\rho$ the likelihood to observe the data $X(t)$ is the same.

\subsection{Predicted Dynamics of Log Price}

We now look into how this expected $X_0(T)$ affects $X$. Integrating by parts the part that is proportional to $\dot X$, we can write
\beq
X_0(T)=(1-\frac{\sigma}{\sigma'})\frac{k\sigma}{\sigma'}\int^T_0dt'\exp\big(-\frac{k\sigma}{\sigma'}(T-t')\big)X(t')+\frac{\sigma}{\sigma'}X(T)+\frac{\sigma}{k\sigma'}(a-\sigma^2/2)+\frac{\sigma'^2}{2k}.
\eeq
In other words, $X_0(T)$ is an weighted average between the exponential moving average (EMA) of the past prices and the current price, with a constant shift proportional to the risk premium of the reasonable price. We can also write the predicted drift of $X$ as
\beq\label{sol_risk_premium}
\mu_X=k(X_0-X)-\sigma'^2/2=k(1-\frac{\sigma}{\sigma'})({\mathrm EMA}-X)+\frac{\sigma}{\sigma'}(a-\sigma^2/2).
\eeq

This is the central result of the paper. We now can easily see what the model predicts. Firstly, the combination $(k\sigma/\sigma')$ defines the scale of the average; the model is trend-following when $\sigma>\sigma'$, and mean-reversing when $\sigma<\sigma'$. When $\sigma=\sigma'$, $X_0$ always differs from $X$ by a constant, and the model is equivalent to a pure geometric Brownian process (i.e., the real price is the reasonable price.) It makes intuitive sense too, as when $\sigma$ is large it means the market price a lot of times is playing catch up, so when there is a trend starting one can expect it is just the beginning of a larger movement. When $\sigma'$ is large, then the market price in the short term is freely wiggling around, but in the longer term has to regress back to where the reasonable price is. 

For a fixed set of parameters, this model thus has a limitation. It is always either trend following or mean reversing, and not flexible enough to dynamically generate regimes that are of different characters. Still, the virtue of the model is to connect the contrast behavior of either trend following or mean reversing to the volatility ratio of the market price and the reasonable price. It is interesting to see in this model, that when we allow $\sigma$ and $\sigma'$ to vary slowly with time, heteroskedasticity naturally leads to trend following and mean reversing regimes.

In Eq. \ref{sol_risk_premium}, we actually only derived the mean of the risk premium $\mu_X$. As $X_0(T)$ is still an unobserved random variable, it has a variance and will contribute to the variance of $\mu_x$. Fortunately in the continuum limit, the variance of $\mu_X$ does not come into play in the evolution of $X$. This is because at every time step $X$ is observed and has no variance. With
\beq\label{sol_x}
d X=\mu_X dt+\sigma' dz',
\eeq
the variance of the change of $X$ (i.e., the log return) is
\beq
{\rm Var}(dX)={\rm Var}(\mu_X) dt^2+\sigma'^2 dt.
\eeq  
As long as the variance of $\mu_X$ is finite (it is in this model) as $dt\rightarrow 0$, the contribution from it is negligible comparing to the contribution from $dz'$. Therefore, if the model assumptions are valid (that is, $\sigma$, $\sigma'$, $k$, and $a$ are really constant in time and both $dz$ and $dz'$ are white noise), the optimal strategy for investing in the instrument will be to hold an amount proportional to $\mu_X(t)/\sigma'$  at any given time.

To conclude this section, we also notice that this model thus gives an theoretical explanation of the concept of ``Bollinger bands"\cite{bollinger} in technical analysis. According to our theory, whether one should buy or sell when the price touches the band edge, depends on the ratio of the two volatilities.

\subsection{Fitting the Model Parameters}
In machine learning, it is common to use either the expectation-maximization technique\cite{EM} or the Viterbi algorithm\cite{viterbi} for model with hidden variables. In our case, however, since the model is integrable, it is  more straightforward just to integrate out $P_0$ and find the maximum likelihood estimates of the model parameters. In fact, due to the model being Gaussian, integrating out $P_0$ is the same as plugging the most probable value (this is actually the Viterbi algorithm in continuum) we have found into the likelihood function.

Nevertheless, we shall not use Eq. \ref{gensol2} directly. The log likelihood function can be diagonalized more easily in Fourier space, where it almost becomes a sum of all different frequencies, if not only because the price movement is not periodic. We will decompose $X$ and $X_0$ into periodic functions and a constant drift in $[0,T]$. The drift is needed, because the derivative term in the log likelihood, when transformed into Fourier space, includes the contribution from the difference of the initial value and the final value. If we use only periodic components (which in principle can faithfully represent any function in $[0,T]$), in effect we are saying that there is a very large change from the final value to the initial value, just before the time series ends. This change is extremely unlikely to happen in the correct model; not including a constant shift will thus distort our fit.

First to simplify the equations a little bit we change variables to $X_0'=X_0+\sigma'^2t/2$ and $X'=X+\sigma'^2t/2$. Then we write
\beq
X_0'=v_0t+\sum_n\mcal{X}_{0n} e^{i\omega_n t}
\eeq
\beq
X'=vt+\sum_n\mcal{X}_ne^{i\omega_nt}
\eeq
with
\beq
\omega_n=\frac{2\pi m}{T};\; m\in \mathbf{Z}
\eeq

We then have
\beq
\int_0^T dt\; \Delta X_0^2= (v_0-b)^2T+T\sum_n\omega_n^2|\mcal{X}_{0n}|^2
\eeq
where $b\equiv a-\sigma^2/2+\sigma'^2/2$.

It is a bit messier for $\Delta X^2$ and $\Delta X\Delta X_0$. It contains cross terms between the constant drift and all Fourier components. We will keep these terms in the following calculation, but  we shall see that when the series we are analyzing are reasonably well described by the model, they will be small. We have

\begin{multline}\label{dx2}
\int^T_0 dt\; \Delta X^2\sim \int^T_0 dt\;\big(v+k(v-v_0)t+k(\mcal X_0-\mcal X_{00})\big)^2\\
+2k(v-v_0)T\sum_{n\neq 0}\big(\mcal X_n+\frac{k}{i\omega_n}(\mcal X_n-\mcal X_{0n})\big)\\
+T\sum_{n> 0}\big(2\omega_n^2|\mcal X_n|^2+2k^2|\mcal X_n-\mcal X_{0n}|^2-2i\omega_n k(\mcal X_n \mcal X_{0n}^*-c.c.)\big);
\end{multline}
\begin{multline}
\int^T_0 dt\; \Delta X\Delta X_0\sim\int^T_0 dt\; (v_0-b)\big(v+k(v-v_0)t+k(\mcal X_0-\mcal X_{00})\big)\\
+k(v-v_0)T\sum_{n\neq 0}\mcal X_{0n}\\
+T\sum_{n > 0}\big(\omega^2_n(\mcal X_{0n} \mcal X_n^*+ c.c)+i\omega_nk(\mcal X_{0n}\mcal X_{n}^*-c.c.)\big).
\end{multline}

First we look at terms that couple with zero frequency variables. Integrating out $v_0$, $\mcal X_{00}$, and $b$, we get
\beq\label{nasty}
(\ln L)_0\propto \frac{6(\sigma (\sum_{n\neq0}k/\i\omega_n(\mcal X_{n}-\mcal X_{0n})  + \sum_{n\neq0}\mcal X_{n}) - \sigma' \rho \sum_{n\neq0}\mcal X_{0n})^2}{\sigma^2\sigma'^2(1-\rho^2)T}.
\eeq

Comparing this term to the remaining finite frequency terms which are proportional to $T$, we see that it is typically very small, when $T$ is large. Specifically, the latter two terms in the numerator are negligible as long as the components do not scale with $T$. This will hold as long as the drift in the series does not vary significantly as a function of time. (If it does, such in the case if we study the time series of a stock passing through the 2008 financial crisis, using this model itself is not going to be a good fit anyway.) When we look at the finite frequency solution of $\mcal X_{0n}$ in below, we will see that $(\mcal X_n-\mcal X_{0n})$ is proportional to $\omega_n$ at small $n$; this means that the first term is also small under the same condition.

The argument above indicates that Eq. \ref{nasty} is small comparing to  all the finite frequency terms. However, it does not say whether the absolute value of this term is much less than one. In the maximum likelihood estimate, the standard deviation of the estimated parameters are usually taken to be the range where the log likelihood drop by $\frac12$. It is therefore necessary to keep terms of order $1$ in the exponential, even if they are small comparing to the rest of the terms. Nevertheless, the argument does imply that the change to the normalization when we integrate over all possible $\mcal X$ will be much less than one when we take the log. It is therefore safe to ignore Eq. \ref{nasty} when calculating the normalization.

When we go back a step before we integrate over $b$, if we ignore the terms coupling to the finite frequency, we get
\beq
(\ln L)_0\propto -\frac{(b-v)^2 T}{2\sigma^2}
\eeq
This implies our maximum likelihood estimate of $b$ is $b=v$, with a standard deviation of ($\sigma/\sqrt{T}$). The terms we ignore causes a systematic error that scales with $(1/T)$, which is much smaller than the standard deviation when $T$ is large.

Now we turn to the finite frequency part of the log likelihood function. First let us ignore terms in Eq. \ref{nasty}. It is then a direct sum of all the frequencies. Maximizing with respect to $\mcal X_{0n}$, we find that it satisfies
\beq
(\sigma'^2\omega_n^2+\sigma^2k^2)\mcal X_{0n}=\big(\sigma^2k^2+i\omega_nk(\sigma^2-\rho\sigma\sigma')+\rho\sigma\sigma'\omega_n^2\big)\mcal X_n.
\eeq
Notice that it agrees with Eq. \ref{x0sol}, as it should. Also one can see that $(\mcal X_{0n}-\mcal X_n)$ is small and proportional to $\omega_n$ at small $n$, as long as $k$ is finite. Now, integrating out $\mcal X_{0n}$, we get
\beq
L_n\propto \exp\bigg(-\frac{T|\mcal X_n|^2(\omega_n^2+k^2)\omega_n^2}{\sigma'^2\omega_n^2+\sigma^2k^2}\bigg);
\eeq 
the $\rho$-dependence is complete cancelled out! Also notice that if $\sigma=\sigma'$, the $k$-dependence is gone, and the likelihood function is the same as a pure random walk with standard deviation $\sigma$.

To get the full likelihood function one only needs to compute the normalization constant in front by integrating out $\mcal X_n$. When there is no interaction between different frequencies this is straightforward to do:
\beq\label{L0}
\log L=\sum_n \log\frac{T(\omega_n^2+k^2)\omega_n^2}{2\pi(\sigma'^2\omega_n^2+\sigma^2k^2)}+\bigg(\frac{T|\mcal X_n|^2(\omega_n^2+k^2)\omega_n^2}{\sigma'^2\omega_n^2+\sigma^2k^2}\bigg).
\eeq

Now we include of Eq. \ref{nasty}. The log likelihood is no longer a direct sum of all the frequencies; in fact, $\mcal X_{0n}$ now is a function of all $\mcal X_n$. We can still numerically solve for $\mcal X_{0n}$, and plug it in the exponential. To calculate the precise normalization with the inclusion of Eq. \ref{nasty}, however, is quite costly when the time series is long (due to the necessity to calculate the log determinant of a $T\times T$ matrix), and we will take advantage of the argument above and approximate it using the original normalization in Eq. \ref{L0}. It is straightforward to compute the log likelihood numerically and maximize it with respect to parameters $\sigma$, $\sigma'$, $k$, $a$, and $\rho$.
\\\\
In this section, we have thus ``solved" the model: given the model parameters and the observed market prices, we can find the best estimate of the theoretical prices in the past, which depends on the market prices both ahead of and after their time, and the latest  theoretical price, which can only depend on the market prices before it. We find that the correlation between the prices does not affect the estimate of the latest theoretical price, and therefore the future market price forecast. From the observed market prices, we can also deduce the most probable values of the parameters $\sigma$, $\sigma'$, $k$, and $a$.

\section{Numerical Test of the Model}
In this section, first we verify our analytic solutions of the model, by applying them on time series that are generated by the model. We then look at real world data and see how this model applies.

\subsection{Verification of Analytic Solutions}
First we generate a time series using the model with some parameters, then we shall find $X_0$ and $X_0(T)$ according to Eq. \ref{gensol2}, and Eq. \ref{x0ind}. There are slight complications due to the fact that our time series is discrete; there is a small difference of the normalization and the exponent of the Green's function, and the precise location of the derivatives on the right hand side needs to be specified. The correct discrete formula can be derived by writing the model explicitly in terms of the discrete variables and their differences. 

One interesting difference between the derivation, here we just mention in passing, is that it does not require integrating by parts to do the ``variation'' in discrete time. The derivative of the price becomes a difference between consecutive prices and can easily be differentiated against the two prices. Eq. \ref{bc}, instead of being a total-derivative integrated to the boundary, comes directly from differentiating against the price on the boundary. In the end, the discrete result will approach the result in the continuum limit when $k\Delta t\ll 1$.

\begin{figure}[h]
\includegraphics[width=15cm]{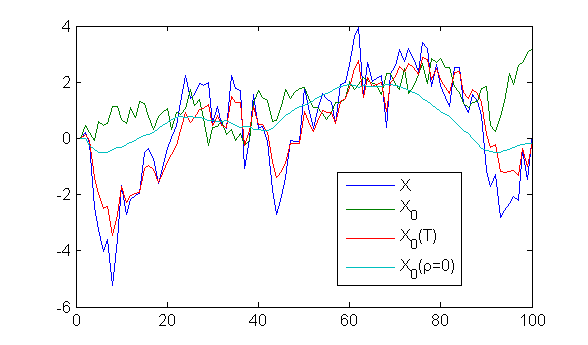}
\caption{Randomly generated market price $X$, with model parameters $\sigma=0.5$, $\sigma'=1$, $k=0.2$, $a=0.125$, and $\rho=0$. $X_0$ in dark green is the underlying reasonable price offset by $-\sigma'^2/2k$, such that the market price is expected to follow it. $X_0(T)$ in red is the most probable reasonable price offset by the same amount if we only know the prices up to that time; $X_0(\rho=0)$ is the most probable reasonable price knowing the whole time series, again offset by $-\sigma'^2/2k$.}
\label{fig1}
\end{figure}

In Figure \ref{fig1}, we plot a randomly generated log price, with $\sigma\sqrt{\Delta t}=0.5$, $\sigma'\sqrt{\Delta t}=1$, $k\Delta t=0.2$, $a\Delta t=0.125$, and $\rho=0$. In the plot we have chosen the unit of time such that $\Delta t=1$. We shall use this unit from now on. As we can see, the blue line does have a tendency to go towards the red line. In this parameter regime, the red line fluctuates smaller than the blue line, and the price is mean-regressing. Notice how we cannot really recover the actual reasonable price, but the most probable reasonable price $X_0(\rho=0)$ at least gets close until the end.

\begin{figure}[h]
\includegraphics[width=15cm]{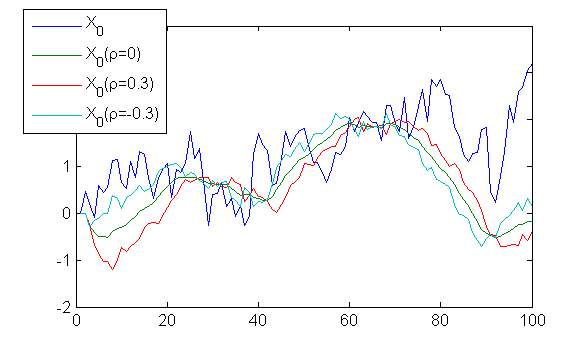}
\caption{The underlying reasonable price $X_0$ and a few maximum likelihood estimate of $X_0$ at various $\rho$. Every series is shifted by the same amount $-\sigma'^2/2k$.}
\label{fig2}
\end{figure}

In Figure \ref{fig2}, we plot several maximum likelihood estimate of $X_0$ at different $\rho$. One can see that at the end of the series they all converge to roughly same value (there are some weak $\rho$ dependence in discrete time actually.)

Now we proceed to fit the model parameters. We shall compare two fits, one includes Eq. \ref{nasty} and the other does not. To start with, we generate $20$ independent time series randomly from the model with $500$ time steps each, using the parameter set $(\sigma,\sigma',k,a,\rho)=(0.05,0.1,0.2,0.002,0.5)$. Then we find out the maximum likelihood estimate of the parameters of either model, and record their mean and standard deviation in the following two tables:
\begin{table}
\centering
\begin{tabular}{|c|c|c|c|}
\hline
&ave&std1&std2\\
\hline
$\sigma$&0.0457&0.0144&0.005\\
\hline
$\sigma'$&0.0965&0.0079&0.0019\\
\hline
$k$&0.1919&0.0864&0.0158\\
\hline
$a$&0.0016&0.0027&0.0022\\
\hline
$\rho$&0.2827&0.4967&0.0238\\
\hline
\end{tabular}\;\;\;\;
\begin{tabular}{|c|c|c|c|}
\hline
&ave&std1&std2\\
\hline
$\sigma$&0.0434&0.0103&0.0041\\
\hline
$\sigma'$&0.0967&0.0046&0.0039\\
\hline
$k$&0.1879&0.0569&0.0122\\
\hline
$a$&0.0015&0.0026&0.0022\\
\hline
$\rho$&0.0241&0.0319&0.0272\\
\hline
\end{tabular}
\caption{The result of maximum likelihood estimates. The estimate on the left takes Eq. \ref{nasty} into consideration whereas the estimate on the right does not.}
\end{table}
In the tables, ``ave" denotes the average of the parameter estimates of the $20$ runs. ``std1" is the sample standard deviation of the $20$ runs, and ``std2'' is the sample average of the standard deviation estimate from the $\chi^2$ analysis where the log maximum likelihood falls by $1/2$. The first table comes from MLE analysis where we include the contribution of Eq. \ref{nasty}, whereas the second table is from when we ignore it.

In general, we see that both analyses give pretty close results, except for the estimation of $\rho$. Excluding Eq. \ref{nasty} wrongly prefers $\rho=0$ for all runs, as one can see by the mean not significantly different from zero, and the smallness of both standard deviations. Including it gives a more erratic behavior of $\rho$; the estimator seems to prefer some arbitrary $\rho$ different for each run, shown by the large std1 and the small std2. In any case, it seems our conclusion in continuum that $\rho$ does not affect the dynamics of $X$ is a reasonable approximation to make, as in either fit the estimated value of $\rho$ is far away from the set value in the model, yet the other parameters are estimated correctly.

Overall, including the contribution of Eq. \ref{nasty} does slightly improve the fit, as seen from the average. However, at the same time its estimate is also more erratic as seen from std1. Still, the MLE from both methods are still biased in the same way and underestimates the two $\sigma$s. In the following subsection, we will set $\rho=0$ and use the MLE without Eq. \ref{nasty} to extract the model parameters. We will report if the two MLEs return significantly different model parameters.

\subsection{Application on S\&P500}
Below is a plot of the S\&P500 index for the past 10 years:\footnote{ The data is downloaded from \href{https://research.stlouisfed.org/fred2/series/SP500/downloaddata}{https://research.stlouisfed.org/fred2/series/SP500/downloaddata}}
\begin{figure}[h]
\centering
\includegraphics[width=10cm]{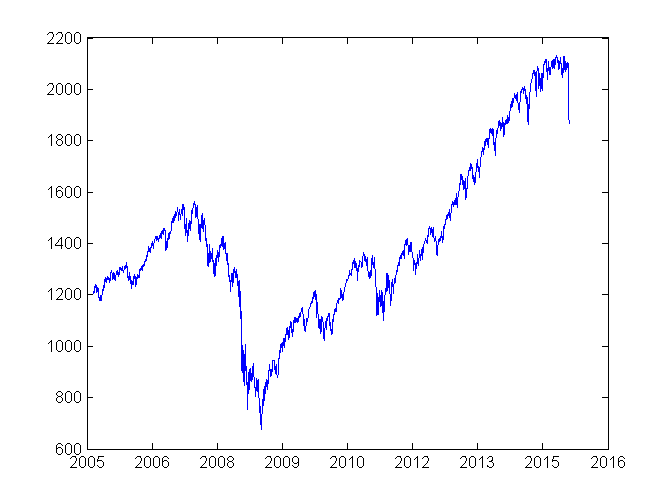}
\caption{S\&P500 index from 2005-08-24 to 2015-08-25.}
\label{fig3}
\end{figure}

\begin{figure}[h]
\includegraphics[width=8cm]{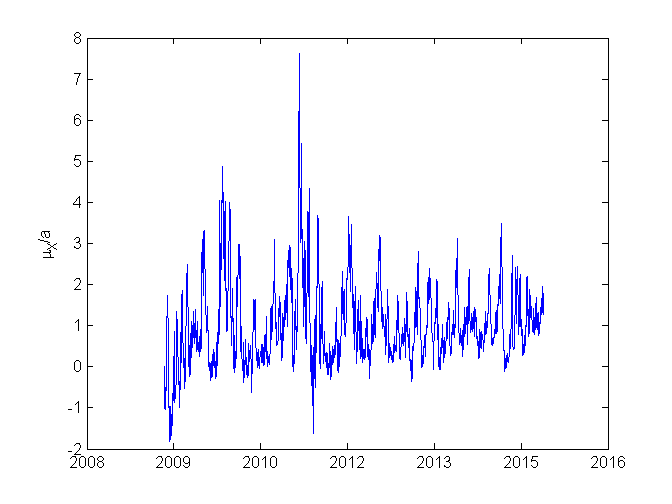}
\includegraphics[width=8cm]{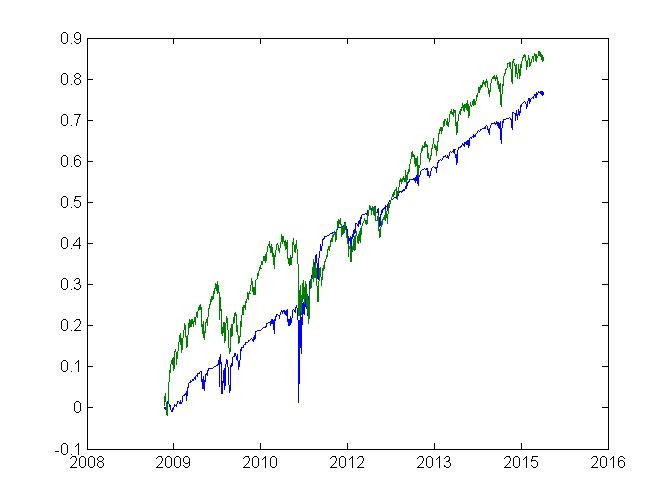}
\caption{(a) The predicted risk premium $\mu_X$ as a function of time. The $y$-axis is in units of the average risk premium $a$. (b) The accumulated return curve. The green curve is holding a constant amount (so it is the same curve as in Fig. \ref{fig3}), and the blue curve is holding an amount proportional to the predicted risk premium. The holding amount is normalized such that the standard deviation of the return is the same as the original return.}
\label{fig4}
\end{figure}

Due to the financial crisis around 2008, it is not a very good idea to fit the whole series with uniform volatility $\sigma$, $\sigma'$ and risk premium $a$. Let us instead start from June 2009, where the market is somewhat recovered, until just before the recent crash. Using the MLE procedure, it results in the following optimal parameters $(\sigma,\sigma',k,a)=(0.0066, 0.0094,    0.0965, 0.0006).$ The fit is an improvement to the geometric Brownian hypothesis if we compare them using the Akaike information criterion (AIC):
\beq
\delta {\rm AIC}=\delta (2k-2\ln L)=4-15.73=-11.73. 
\eeq
In the equation, $k=2$ is the difference of fitting parameters in the two models, and we have calculated the log likelihood for them. (Conveniently the geometric Brownian motion $dX/X=adt+\sigma dz $ with parameters $(\sigma,a)$ is given by the parameters $(\sigma,\sigma,0,a)$ in our model.) We can also check the model is working, by regressing its predicted risk premium against the actual log return:
\beq
	r_i\equiv (\ln P_{i+1}-\ln P_i)\sim\mu_{Xi}.
\eeq
We get
\begin{verbatim}
Estimated Coefficients:
                   Estimate       SE            tStat       pValue   
    (Intercept)    -0.00014227    0.00034876    -0.40794      0.68337
    x1                  1.2212       0.43898      2.7818    0.0054706


Number of observations: 1560, Error degrees of freedom: 1558
Root Mean Squared Error: 0.00972
R-squared: 0.00494,  Adjusted R-Squared 0.0043
F-statistic vs. constant model: 7.74, p-value = 0.00547
\end{verbatim}
This is to be compared with our theoretical prediction from the model
\beq
dX=\mu_X dt+\sigma' dz
\eeq 
such that the intercept is zero, the slope $x1=1$ and $R^2={\rm Var}(\mu_X)/({\rm Var}(\mu_X)+\sigma'^2)=0.0036$. (Notice that this sample variance of $\mu_X$ is different from the variance of the unobserved random variable $X_0(T)$ we talked about in section 3.2.) We plot the obtained risk premium $\mu_X$ as a function of time in Fig. \ref{fig4}a.

If the model assumptions are correct, the optimum strategy to invest is to hold an amount proportional to the risk premium. However, doing so results in the earning curve shown in Fig. \ref{fig4}b: with the standard deviation normalized to be the same, we see that unlike predicted in theory such that the Sharpe ratio should increase by a factor of $\sqrt{E[\mu_x^2/a^2]}\sim\sqrt 2$, it actually decreases by 10\%. 

The reason for this investing strategy to be ineffective, given the fact that the predicted risk premium does predict the return within model specification, is the following. First, the volatility of the residual return, as can be seen evidently from the curve, is not a constant. Even if the risk premium predicted from our model ends up to be somewhat accurate, in the strategy the holding amount now needs to be divided by the time-varying standard deviation. Second, one can check that the residual returns are not completely independent with one another, unlike assumed in the model. With such correlation, the holding amount needs to be multiplied by the inverse of the correlation matrix.

When the homoskedasticity and the independent conditions are satisfied, the strategy will perform more satisfactorily. In fact, if we focus on the later half of the curve from 2012 to 2015, the overall Sharpe ratio for the strategy is 75.9, whereas the Sharpe ratio for holding a constant amount is 62.1.

\begin{figure}[h]
\centering
\includegraphics[width=12cm]{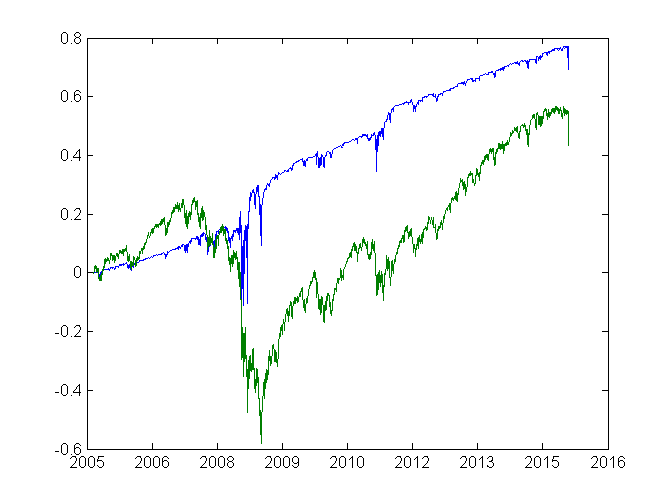}
\caption{Accumulated return from 2005 up to now, using the strategy. The green curve is holding constant. Again the two curves are scaled such that the returns have identical sample standard deviations.}
\label{fig5}
\end{figure}

We conclude this section by consider using the strategy starting from 2005. The accumulated return is shown in Fig. \ref{fig5}. There is no theoretical foundation behind, but the strategy seems to work pretty well even during the market collapse!

\section{Discussion}
In this paper, we have proposed a model such that there is an unobserved reasonable price following geometric Brownian motion with the reasonable risk premium, and the market price tries to follow it. Our main result is that the market price will be trend following, if the standard deviation of the market price is smaller than of the reasonable price, and mean-reversing if it is the other way around. We have also developed the MLE to estimate the model parameters from a given time series. We have shown that from 2009 upto now the S\&P500 index can be described better by the model than a pure geometric Brownian motion, and it is in the mean-reversing regime.

There are a few technical issues regarding the difference between the continuous time and the discrete time formulation, which I deliberately tried not to mention above. 

One problem that is central to our result is whether $\rho$ is really not measurable and have no implications. In the continuum it certainly is, but from my numerical results it does not seem to be entirely true in discrete time. The MLE results all prefer some $\rho$ but the preference seems unrelated to the value set when the time series is generated. The most likely explanation is that this preference is a result of the approximation we have done, either when we ignore part or the entirety of Eq. \ref{nasty} or when we ignore its contribution to the overall normalization. If everything is done accurately, there should still be preference of $\rho$, but the preference should not scale with the length of the time series.

Then there is the question of how the discrete time series approach the continuous time. The answer will be that it solely depends on the value of the parameter $k$ in the discrete formulation. If we view the discrete time series as an approximation to the continuous time, the parameter $k$ is $kdt$ in the continuous time formulation. The analytical results in this paper thus are only valid when $k\ll 1$. One notable example is that in addition to the mean given by Eq. \ref{sol_risk_premium}, the variance of $\mu_X$ will also need to be considered when $k$ is not small.

One question arises when we think about predicting in-sample returns using the predicted risk premium. The question is, does our MLE procedure minimize the error of this prediction, such that it selects the model parameters that produce the minimal error of predicted return? It is not obvious at all from the calculation in section 3.3, especially since the likelihood function contains two errors, one from $X$ and one from $X_0$. The integration over $X_0$ does not eliminate the contributions from $\Delta X_0$, and plugging in the most probable value of $X_0(t)$ in $\Delta X$ does not give the correct error from prediction, since $X_0(t)$ is in general different from $X_0(T)$, which is what we plug in when we calculate the risk premium.

However, our conjecture to this question is that the MLE does minimize the error of the predicted return in the continuum limit. This stems from the fact that in continuum the evolution of $X$ is given by Eq. \ref{sol_x}, where we can just replace $\mu_X$ by its mean value. This necessarily means the likelihood function of the original model integrated over $X_0$ should be the same as calculated from Eq. \ref{sol_x}, which is proportional to $\exp(\int dt \epsilon^2/2\sigma'^2)$. 

When the discrete nature of the series become apparent, the randomness of $X_0(T)$ then cannot be ignored in Eq. \ref{sol_x}. The error of the predicted return becomes a sum of two parts: first is the difference between the realized return and $X_0(T)$ and the second is the difference between the mean of $X_0(T)$ and the actual realization. The two parts are optimized with different coefficients and it is unclear such optimization will result in their sum squared being minimized.

One interesting observation is that in the discrete form the model can be rearranged to look very similar to a ARMA(2,1) model with some fixed relations among its parameters; one big difference still is there are two random sources in the model and we are not sure if it can somehow be transformed into one.

Finally, we throw out some ideas how this model can be realistically used. The immediate improvement is to include the possibility that the standard deviations can change with time. In the minimal extension it might be enough to add a few discrete variables, each describes a state with the model with a different set of parameters. switching between the states can be some additional Markovian dynamics, or just make them hidden and determine the current state by some statistic. In practice we can also treat the predicted risk premium as some general indicator and use them along with past return to form some generalized AR(n) models to eliminate the correlations among errors of predicted returns. It would also be interesting to see whether this model can characterize dynamics at a much lower time scale, such as intraday movements in minutes or seconds.

It would also be interesting to extend this model to include multiple instruments (whether each instrument errands from the covariance with the pricing portfolio on mean-variance front, or the pricing portfolio itself errands in its composition or price, or something in between) or to consider derivative pricing, such as options, when the underlying instrument follows the process prescribed by this model.

\end{document}